# Quantum Critical Behavior in a Concentrated Ternary Solid Solution

Brian C. Sales[*], Ke Jin, Hongbin Bei, G. Malcolm Stocks, German D. Samolyuk, Andrew F. May, Michael A. McGuire

Materials Science and Technology Division, Oak Ridge National Laboratory

**\*Correspondence to:** salesbc@ornl.gov

*This manuscript has been authored by UT-Battelle, LLC under Contract No. DE-AC05-00OR22725 with the U.S. Department of Energy. The United States Government retains and the publisher, by accepting the article for publication, acknowledges that the United States Government retains a non-exclusive, paid-up, irrevocable, world-wide license to publish or reproduce the published form of this manuscript, or allow others to do so, for United States Government purposes. The Department of Energy will provide public access to these results of federally sponsored research in accordance with the DOE Public Access*

# Quantum Critical Behavior in a Concentrated Ternary Solid Solution

Brian C. Sales, Ke Jin, Hongbin Bei, G. Malcolm Stocks, German D. Samolyuk, Andrew F. May, Michael A. McGuire

Materials Science and Technology Division, Oak Ridge National Laboratory

*Correspondence to: salesbc@ornl.gov

**Quantum critical behavior has been associated with some of the most exotic emergent states of matter including high-temperature superconductivity.[1,2] Much of the research into quantum critical point (QCP) physics has been hampered by the lack of model systems simple enough to be analyzed by theory.[3] Here, we show that the concentrated solid solution fcc alloys, including the so-called high-entropy alloys, are *ideal model systems* to study the effects of chemical disorder on emergent properties near a quantum critical region. The face centered cubic (fcc) alloy $NiCoCr_x$ with $x \approx 1$ is found to be close to the Cr concentration where the ferromagnetic transition temperature, $T_c$, goes to 0. Near this composition these alloys exhibit a resistivity linear in temperature to 2 K, a linear magnetoresistance, an excess $-TlnT$ contribution to the low temperature heat capacity and excess low temperature entropy. All of the low temperature electrical, magnetic and thermodynamic properties of the alloys with compositions near $x \approx 1$ are not typical of a Fermi liquid and suggest strong magnetic fluctuations associated with a quantum critical region. The limit of extreme chemical disorder in these simple fcc materials thus provides a novel and unique platform to study quantum critical behavior in a highly tunable system.**

The recent synthesis of single-phase concentrated solid solutions has provided new frontiers for material physics research.[4,5] These alloys exhibit exceptional mechanical properties as compared to conventional alloys[6] and exhibit enhanced resistance to radiation damage[7]. The alloys are characterized by a simple face-centered cubic (fcc) or body-centered cubic crystal structure (bcc) and two or more principle elements. If the alloys have five or more elements in equal atomic ratios (e.g. NiCoCrFeMn), they are often referred to as high-entropy alloys[7]. Most traditional multielement alloys have multiple phases and complex microstructures, or often have a majority element that can be considered the solvent and minority elements as dopants or solutes. In single-phase concentrated solid solutions, however, the elements randomly occupy an ordered fcc or bcc lattice with all elements having roughly equal concentrations. This results in locally disordered chemical environments, unique site-to-site distortions, and extreme complexity at the electron level.  In spite of the inherent chemical disorder, these alloys are single phase with a simple cubic crystal structure and can be prepared as large single crystals. The alloys thus represent an interesting class of model materials for investigating the effects of a particular type of disorder on the cooperative response of the electrons, spins and lattice.

During our investigation[8] of the basic electrical, magnetic, and thermal transport properties of a series of eight fcc Ni-based single-phase single-crystal concentrated alloys (Ni, NiCo, NiFe, NiFeCo, NiCoCr, NiCoCrFe, NiCoCrFeMn, NiCoCrFePd), we noticed that the low temperature resistivity of the NiCoCr alloy was linear in temperature at least down to our base temperature of 2 K. This is highly unusual for a metallic alloy in this temperature regime and was not observed for any of the other Ni-based alloys, which exhibited a resistivity that followed the expected $T^2$ or $T^5$ behavior.[8] A linear resistivity for a metal at low temperatures, although not fully understood, is often an indication of the proximity to an unusual quantum state associated with a quantum critical point (QCP) .[9] As shown below this is indeed the case for NiCoCr$_x$ which is close to a ferromagnetic QCP for x ≈1. Simple fcc single-phase alloys with compositions near NiCoCr thus appear to be model systems for investigating how chemical disorder affects the cooperative response of the charge, spin, and lattice degrees of freedom near a QCP (or a quantum critical region as is the case for systems with significant disorder)[10,11]. Characterizing and understanding the physical signatures of QCP behavior in this new class of alloys provides the motivation for the present work. The concentrated NiCoCr$_x$ alloys share some features of quantum critical behavior with other systems at low temperatures, such as a linear resistivity at very low temperatures (2K), an additional $\approx -T\ln T$ contribution to the low temperature heat capacity, and a linear magnetoresistance. However, an exponential variation of the dc magnetic response with concentration was observed, which was totally unexpected.

The resistivity of NiCoCr from 2–300 K is shown in the inset of Fig 1. The residual resistivity, $\rho_0$, is high in this family of concentrated alloys, typically in the range of 90 $\mu\Omega$-cm. This high value can be understood quantitatively using the *ab initio* Korringa-Kohn-Rostoker Coherent-Potential-Approximation (KKR-CPA) theoretical formalism as has been explained previously[7,8] and is briefly discussed again in the supplemental information. The addition of Cr to a ferromagnetic NiCo alloy disrupts electron transport in both the minority and majority spin channels, resulting in a rapid increase in the residual resistivity. This formalism also nicely explains the rapid decrease in the Seebeck coefficient from about 30 $\mu$V/K for NiCo at 300 K to 0.9 $\mu$V/K for NiCoCr. The addition of Cr also leads to a type of frustration since the spins on the Cr atoms want to be antiparallel to neighboring Cr atoms and to the Ni and Co spins, which cannot be perfectly satisfied in a random solid solution on an fcc lattice. In insulating spin systems frustration can lead to a spin liquid that is characterized by strong quantum fluctuations and no magnetic order to T=0[12]. The main panel of Fig. 1. illustrates the linearity of the resistivity of NiCoCr between 2 and 50 K, which is very unusual for a metal in this temperature range.[9] For NiCoCr$_x$ alloys well away from x =1, a more normal metallic resistivity is recovered.

The response of the resistivity of NiCoCr to a magnetic field is shown in Fig 1b. As the magnetic field is increased, the low temperature linear resistivity is lost (Fig. 1b) and the temperature dependence resembles a normal metal. The data look qualitatively similar to what occurs in $Sr_3Ru_2O_7$, except for that compound the resistivity is linear with an applied field $\mu_0H$= 7.9 T, and more "normal" at other magnetic fields (including H= 0)[9]. The low temperature magnetoresistance, although small in magnitude ($\approx$ 0.5% at 10 T) is remarkably linear at 2 K with no evidence of saturation (Fig. 1c). A linear magnetoresistance is found in graphene, topological insulators, and some other quasi-2D bulk materials where it is associated with linear bands near the Fermi energy and Dirac fermions and nodes[15,16]. However a linear magnetoresistance can also occur in simple metals such as potassium and copper.[17,18] For NiCoCr, the origin of the linear transverse magnetoresistance is not clear although the combination of a high residual resistivity and metallic carrier concentration ($\approx 9 \times 10^{22}$ electrons/cm$^3$ estimated from Hall data- see supplemental) should result in a small magnetoresistance, as is observed.[18] In addition, a linear positive transverse magnetoresistance is observed at 2 K for all of the NiCoCr$_x$ alloys (including x=0), which suggests it is related to common features of the electronic band structure and chemical disorder (see supplemental) and not related to the quantum critical region.

The magnetic susceptibility and magnetization data for NiCoCr are shown in Fig. 2. The susceptibility data are best described as an enhanced Pauli paramagnet, similar in magnitude to Pd metal[19], with no evidence of magnetic order down to 2 K. The magnetization curve at 300 K is linear up to our highest measuring field (5 Tesla), but at 5 K there is a small amount of curvature (Fig. 2 inset). The extrapolation of a linear fit to the high field magnetization data (3-5 Tesla) to $\mu_0 H = 0$ yields a moment of 0.001 $\mu_B$/atom. If NiCoCr were ferromagnetic, this would be an estimate of the spontaneous moment.

The origin of the unusual properties exhibited by NiCoCr is investigated in a series of alloys, NiCoCr$_x$, where the Cr concentration is varied with 0 < x < 1.2. All of the alloys have the same simple fcc structure with the three elements randomly distributed on the fcc lattice (see supplemental section for more details). As previously noted in Fig. 2, for values of x outside of the critical region the resistivity is no longer linear at low temperatures. Qualitatively, the effect of a magnetic field on the resistivity of NiCoCr, is similar to the effect of reducing x. (Compare Fig 1b., 8T data with Fig. 1a, x=0.6) although there is no obvious scaling procedure.

The evolution of ferromagnetism in alloys with reduced Cr concentration is particularly striking (Fig. 3). Both the ferromagnetic transition temperature, $T_c$, and the low temperature spontaneous moment per atom, $M_0$, change exponentially with

x. The estimated Curie temperatures obtained from Fig. 3 (and Arrott plots- see supplemental) are 212, 77, 20, and 2.5 K for x= 0.5, 0.6, 0.7 and 0.8 respectively. The remarkably rapid variation of $T_c$ and $M_0$ with x was not expected by a KKR-CPA mean field theory calculation (Fig. 3c) or by comparing the NiCoCr$_x$ to other magnetic QCP systems such as Ni$_x$Pd$_{1-x}$[20], Cr$_{1-x}$V$_x$[21], Ni$_{1-x}$Cr$_x$[22] where the concentration dependence of $T_c$ on x is usually linear or close to linear. The strong and rapid suppression of ferromagnetism for 0.5 < x < 1.0 suggests that unusually strong magnetic fluctuations and frustration prevent magnetic order. In addition, the small values of the average magnetic moment per atom (Fig. 3) should make these alloys more susceptible to quantum effects. [12]

In spite of the extremely small saturation moments shown in Fig. 3, the magnetic susceptibility data well above $T_c$ are described by a Curie -Weiss law (see Supplemental information for more details) with relatively large effective magnetic moments per atom as estimated from the Curie constant; $p_{eff}$ = 0.8, 1.3, 2, and ≈2.2$\mu_B$/atom for x= 0.8, 0.7, 0.6, 0.5. This is often the case for highly itinerant magnets such as ZrZn$_2$.[23] A phenomenological measure of itinerant ferromagnetism was proposed by Rhodes and Wohlfarth[24] who found that larger values of $q_c/M_o$, termed the Rhodes-Wohlfarth (RW) parameter, corresponded to a more itinerant ferromagnet, where $p_{eff}^2 = q_c(q_c+2)$. The RW parameters for the x= 0.8, 0.7, 0.6, and 0.5 alloys are 19, 12, 8.5, and ≈5. These large RW values are consistent with the presence of strong spin fluctuations.[25,26] A consequence of strongly itinerant ferromagnetism is that $p_{eff}$ is determined by microscopic parameters of the band structure, and should not be regarded as implying a large local atomic magnetic moment.[26]

Low temperature heat capacity measurements can probe some aspects of low energy magnetic or electronic excitations in a temperature regime where the contribution of the lattice is minimal. These have been frequently used to characterize, for example, non-Fermi liquid behavior near a magnetic instability in heavy fermion compounds like CeCu$_{5.9}$Au$_{0.1}$[27,28]. The heat capacity data from a series of NiCoCr$_x$ alloys with 0.5 < x < 1.2 are shown in Fig 6. Above 30 K the heat capacity, C, (Fig. 4a) of all of the alloys are the same within experimental error, which is approximately the size of the data points. Although magnetization data indicates magnetic order for alloys with x= 0.8, 0.7, 0.6 and 0.5, no anomaly is evident in the heat capacity data near $T_c$, presumably because of the very small saturation moment (see Fig. 3). A similar behavior was found for the weak ferromagnet ZrZn$_2$ with $T_c \approx 28$ K and $M_0 \approx 0.16$ $\mu_B$ per formula unit.[29] The heat capacity of the NiCoCr$_x$ alloys are quite different below 20 K. The low temperature heat capacity data from the Pauli paramagnet NiCoCr$_{1.2}$ appears to be a reasonable reference baseline for the other alloys. Standard C/T versus $T^2$ of the NiCoCr$_{1.2}$ data are linear below 10 K yielding a Debye temperature, $\Theta_D$, of 466 K and an electronic specific heat coefficient, $\gamma$, of 9.2 mJ K$^{-2}$ mole-atoms$^{-1}$. Both of these values are reasonable for this type of transition metal alloy.[19] As the Cr concentration is reduced below x = 1.2, there is a monotonic increase in C/T up to x = 0.7. For x= 0.7 and x=0.8 C/T *increases* with cooling below 7 K. As x is reduced further away from the critical region toward the more ferromagnetic alloys (x = 0.6, 0.5), the low temperature C/T decreases in magnitude and begins to return to more normal metallic behavior. The excess low temperature entropy, $\Delta S$, of all of the alloys with respect to the NiCoCr$_{1.2}$ alloy, is estimated by integrating C/T from 0 to 30 K. A linear extrapolation of the C/T data was used below 2 K. The variation of $\Delta S$ with x is shown in the inset of Fig. 4b. The maximum in $\Delta S$ of about 0.9 J K$^{-1}$ mole-atoms$^{-1}$, occurs for x $\approx$ 0.65. The excess entropy at low temperatures is presumably associated with magnetic fluctuations. The amount of excess entropy is substantial when compared to the total magnetic entropy of a mole of spin ½ particles= Rln2 = 5.76 J K$^{-1}$ mole-atoms$^{-1}$. In Fig. 4c part of the data shown in 4b are plotted versus log T. Although we can only measure down to 2 K, the data below 5 K for x= 0.7 and 0.8 appears to be linear in lnT. A $-TlnT$ contribution to the heat capacity has been reported by Lohneysen[27,28] near the QCP of a heavy fermion alloy, CeCu$_{5.9}$Au$_{0.1}$, from 0.1 to 4 K and may be a common feature in many heavy fermion systems displaying non-Fermi-liquid behavior.[27,28] We note, however, that a power law (C $\approx$ T$^\alpha$), which has been proposed for some systems with strong disorder near a QCP (sometimes called Griffiths' singularities), would also describe the low temperature heat capacity data for x= 0.7 and 0.8[10,11,30] down to 2 K.

We have shown that concentrated solid solution transition metal alloys (NiCoCr$_x$) with a simple fcc structure display many of the transport, magnetic and thermodynamic signatures exhibited by more structurally complex compounds near a QCP [27,10], such as a linear resistivity at low temperatures and a –TlnT contribution to the heat capacity. The exponential dependence of the Curie temperature and saturation moment on composition near the QCP is perhaps the most striking and unique characteristic of these alloys. This class of alloys should provide a gateway to a clearer understanding of the role of disorder on the general behavior of matter near a ferromagnetic quantum critical point. Since these materials can be prepared as very large single crystals, powerful techniques such as inelastic neutron scattering and nuclear magnetic resonance can be applied to probe the dynamics of the electrons, spins, and lattice in these model materials.

Although mean field theories, such as KKR-CPA, capture much of the basic physics of these alloys, there are clearly large deviations between theory and experiment (e.g. see Fig 3c) at low temperatures in the vicinity of the quantum critical region. It is hoped that the results reported in this letter will stimulate the development of theories and scaling relationships similar to those proposed to explain the physics of the f-electron compounds near a QCP. [28,30]

*Methods*

Polycrystalline samples of NiCoCr$_x$ are prepared by arc-melting appropriate mixtures of the elements in an argon atmosphere. Each sample is melted and flipped a minimum of five times to insure complete mixing of the three elements. The well-mixed alloy is then dropcast into a 2 mm diameter rod, which is then polished to the desired shape. For resistivity measurements each alloy is polished to a bar with typical dimensions 10 x 0.3 x 0.3 mm$^3$ or for Hall measurements to about 10 x 1 x 0.15 mm$^3$. Single crystals of NiCoCr and NiCo are grown from the well-mixed polycrystalline alloys using a floating zone furnace. We found that the magnetic, transport and heat capacity data are virtually identical between the polycrystalline and single crystal samples of NiCoCr (see supplemental) and hence only data from polycrystalline samples are reported in the manuscript. Careful examinations of polished sections of each alloy using scanning electron microscopy and energy dispersive x-rays confirmed a complete random solid solution, as found by previous work.[6] Resistivity, Hall and heat capacity data are collected using a Physical Property Measurement System (PPMS) using standard methods. Four to six electrical contacts (0.05 mm Pt wires) are spot-welded to each alloy. Magnetic measurements are made with a Magnetic Property Measurement System (MPMS). See Supplemental Information for more details.

**Acknowledgements**

This research was supported primarily by the Department of Energy, Office of Science, Basic Energy Sciences, Materials Sciences and Engineering Division (B.C.S, A. F. M., M. A. M.). K.J., H. B., G. D. S., and G. M. S. were supported by the Energy Dissipation to Defect Evolution (EDDE), an Energy Frontier Research Center funded by the U. S. Department of Energy, Office of Science, BES.

**Author contributions**

B. C. S. conceived and coordinated the research, made all of the polycrystalline samples, and most of the resistivity, magnetic and heat capacity measurements. K. J. and H. B. synthesized well-characterized single crystals of NiCoCr and NiCo and determined the microstructure of the polycrystalline samples. A. F. M. performed Hall measurements and analysis, and M. A. M. provided lattice constant data and analysis. G. M. S. and G. D. S. provided first principles electronic structure calculations. B. C. S. wrote the paper with input from all of the authors.

**Addition information**

Supplemental information is available with the online version of this article.

**Competing financial interests**

The authors declare no competing financial interests.

**FIGURE CAPTIONS**

Fig. 1. **Linear Resistivity and Magnetoresitance** (a) Resistivity, $\rho$, of NiCoCr versus temperature from 2 K to 300 K (inset). The main panel shows the linearity of ($\rho$-$\rho_0$) for NiCoCr versus temperature, where $\rho_0$ is the residual resistivity. A linear fit to through the ($\rho$-$\rho_0$) data is also shown. Also shown in main panel is ($\rho$-$\rho_0$) versus temperature for two compositions away from the critical region. NiCoCr$_{1.2}$ is a Pauli paramagnet, and NiCoCr$_{0.6}$ is a ferromagnet with T$_c$ ≈75 K. For both compositions the temperature dependence of the resistivity deviates strongly from linear. The increase in the resistivity below 5 K for NiCoCr$_{0.6}$ may be due to the Kondo effect but is more likely due to the effects of electron-electron scattering in a disordered system, which can produce similar behavior. [14,15] (b) Resistivity of NiCoCr versus temperature for applied magnetic fields of $\mu_0$H = 0, 1,2,4,and 8 Tesla. (c) Magnetoresistance of NiCoCr at 2K illustrating the linear dependence on magnetic field. Since all of the NiCoCr$_x$ alloys (including x = 0) exhibit a similar linear magnetoresistance (see Supplemental section), this behavior is not associated with the quantum critical region and is likely directly related to the effects of disorder.

Fig. 2 **Magnetic susceptibility versus temperature for NiCoCr** with an applied magnetic field of 1 Tesla. There is no evidence of magnetic order and the magnitude of the susceptibility is similar to that of Pd metal. Magnetization curves are linear at 300 K but show a small amount of curvature at 5 K (Inset).

Fig. 3 **Magnetization data-experiment and theory** (a) Magnetization versus temperature in a small applied magnetic field of 0.01 T for four NiCoCr$_x$ alloys. (b) Magnetization vs magnetic field at T = 5 K for the same four alloys. (c) Estimate of the spontaneous magnetic moment versus Cr concentration from magnetization curves at 5 K (see Fig. 3b and Fig. 2). M$_0$ is determined from linear fits to magnetization curve data for magnetic fields between 3

and 5 Tesla extrapolated back to $\mu_0 H = 0$ (see Fig 2 inset). The red line is an exponential fit to the $M_0$ data [$M_0 = 226*\exp(-12.36 x)$]. The inset shows the dependence of $T_c$ on x, which has a similar exponential dependence [$T_c = 4.0019 \times 10^5 *\exp(-14.64 x)$]. From KKR CPA calculations, the average spontaneous moment per atom should vary linearly with x. The blue line is a linear fit to the calculated values for $M_0$. For x = 0 (not shown) and 0.5, which is far away from the critical Cr concentration region of $x_c \approx 0.7$-1, the theoretical value of $M_0$ is close to the measured value, but the values diverge as $x_c$ is approached. For x = 0, the experimental and theoretical values for $M_0$ are 1.34 and 1.11 $\mu_B$/atom respectively.

Fig. 4. **Heat capacity data** (a) Heat capacity divided by temperature (C/T) versus temperature for NiCoCr$_x$ alloys for 0.5 < x < 1.2. (b) Same data as shown in (a) for T < 30 K. Inset shows the excess entropy, $\Delta S$, versus Cr concentration x. $\Delta S$ is calculated with respect to the x = 1.2 heat capacity data. The maximum excess entropy occurs for x ≈ 0.65 and is about 15% of Rln2. (c) C/T data versus log T for 0.7 < x < 1.2. For x= 0.7, 0.8, the C/T data below 5 K are linear in -ln T down to our lowest measuring temperature of 2 K.

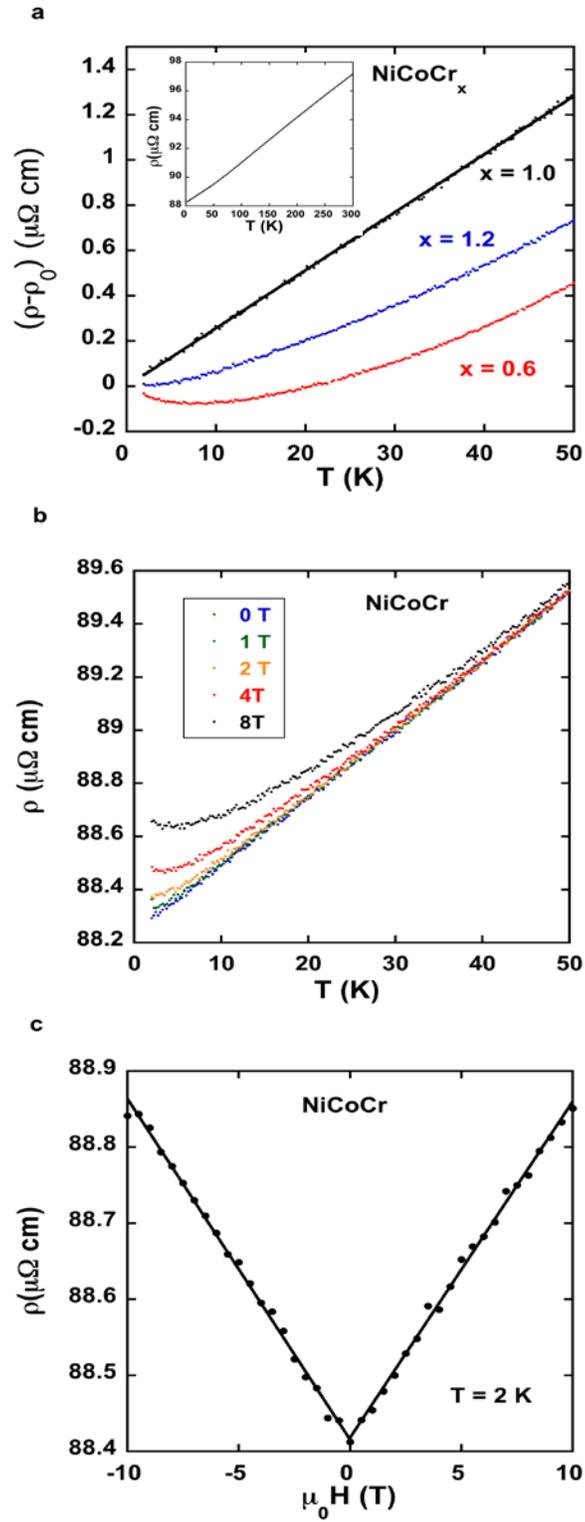

**Figure 1**

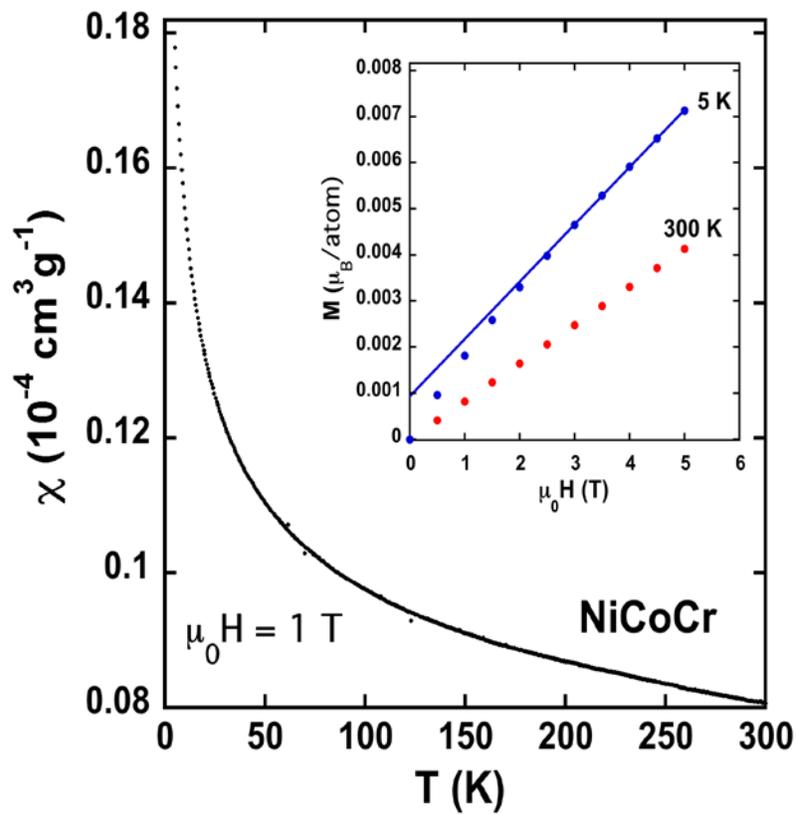

**Figure 2**

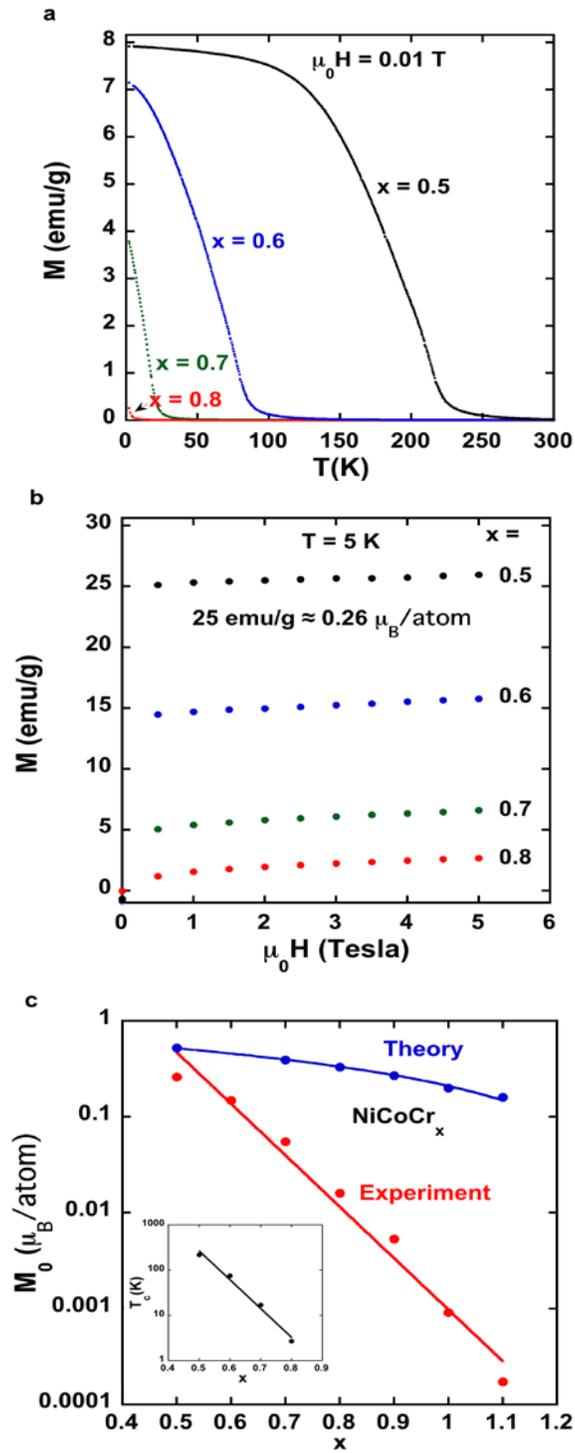

**Figure 3**

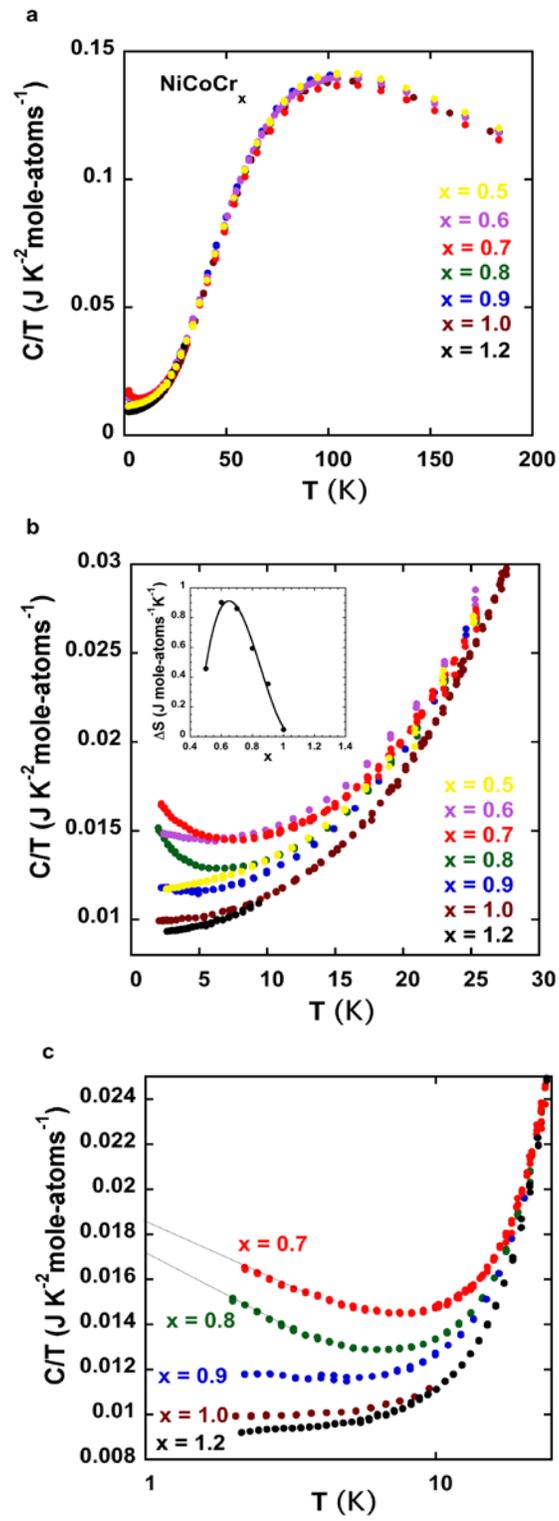

**Figure 4**

*Supplementary Information*

**Quantum Critical Behavior in a Concentrated Ternary Solid Solution**

Brian C. Sales, Ke Jin, Hongbin Bei, G. Malcolm Stocks, German D. Samolyuk, Andrew F. May, Michael A. McGuire

*Materials Science and Technology Division, Oak Ridge National Laboratory*

**S1 Lattice Constants versus Composition**

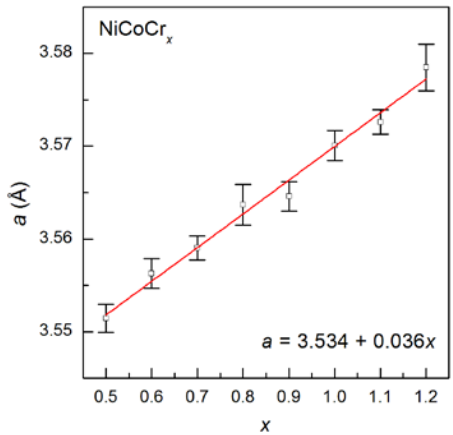

**Figure S1.** Lattice constants vs. Cr concentration x are determined from x-ray diffraction data from a flat polycrystalline surface. A PANalytical X'Pert Pro X-ray diffractometer with monochromatic Cu K$\alpha_1$ radiation was used for the measurements.

**S2 Transport, magnetic and heat capacity measurements**

Four wire resistivity measurements are made on rectangular bars of each alloy with approximate dimensions of 10 x 0.3 x 0.3 mm³, using the resistivity option (AC mode, I= 4mA) of the Physical Property Measurement System (PPMS) from Quantum Design. Platinum wires with a diameter of 0.05 mm are spot-welded onto each sample. The contact resistance for all leads is less than 0.2 Ohms. Transverse magnetoresitance measurements are made on the same bars with the magnetic field perpendicular to the current I. Hall measurements were made on samples with typical dimensions of 10 x 1 x 0.15 mm³ using four spot-welded leads in the standard Hall geometry.[1] Even with applied fields of up to 10 T, the Hall voltage is small for these alloys, yielding an approximate carrier concentration of ≈9 x 10²² holes/cm³.

Magnetic data are taken with the Magnetic Property Measurement System (MPMS) from Quantum Design, which is a SQUID magnetometer. Magnetic moment data are collected as a function of temperature (2-400 K) in fixed applied fields of 0.01 to 1 Tesla, using both field cooled (FC) and zero-field-cooled (ZFC) protocols. Magnetic moment data (magnetization curves) are also collected at fixed temperatures with applied fields between 0 and 5 Tesla.

Heat capacity data are collected with the PPMS heat capacity option using the conventional relaxation method employing small heat pulses with the sample temperature rise restricted to 2% of the sample temperature.

## S3 Average grain size and chemical homogeneity

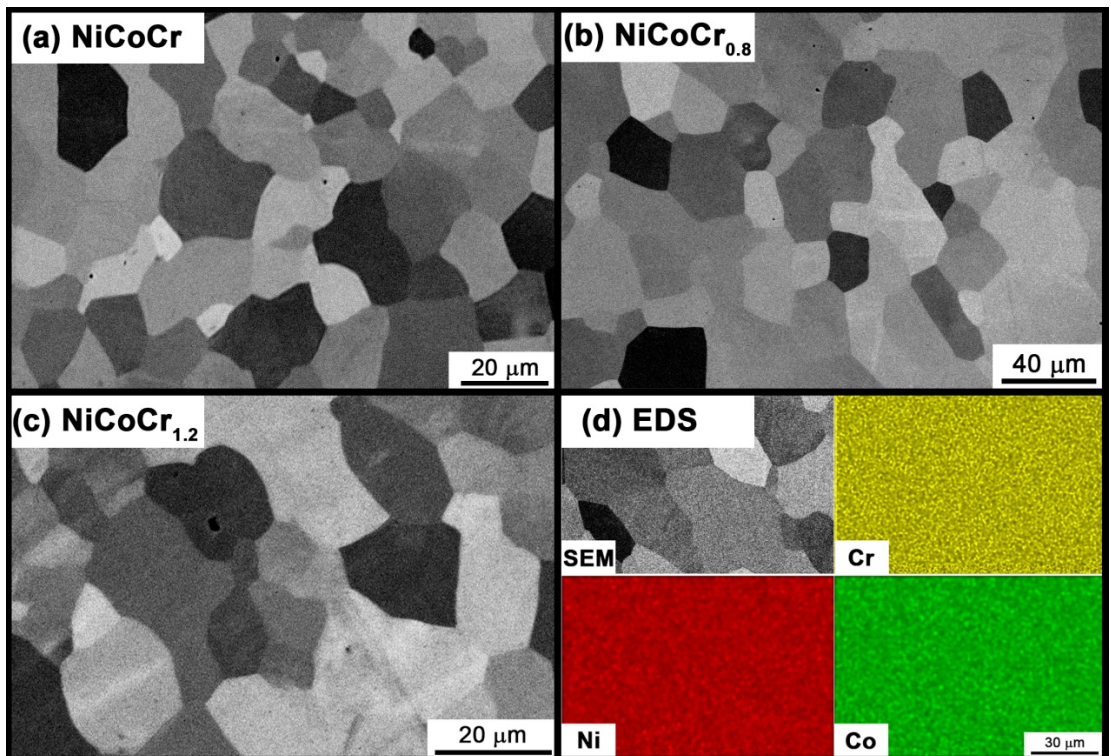

**Figure S2.** Microstructures and composition mapping of polycrystalline NiCoCr$_X$ alloys. (a-c) Electron backscatter images show the grain structures of NiCoCr, NiCoCr$_{0.8}$ and NiCoCr$_{1.2}$, respectively. (d) EDS mapping shows the homogeneous distributions of elemental Cr, Ni and Co in the NiCoCr polycrystalline alloys. All three alloys exhibit roughly equiaxed grains, with average grain sizes of 12 μm for NiCoCr, 16 μm for NiCoCr$_{1.2}$ and 26 μm for NiCoCr$_{0.8}$, respectively.

## S4. Comparison of Polycrystalline NiCoCr vs Single Crystal NiCoCr Transport Data

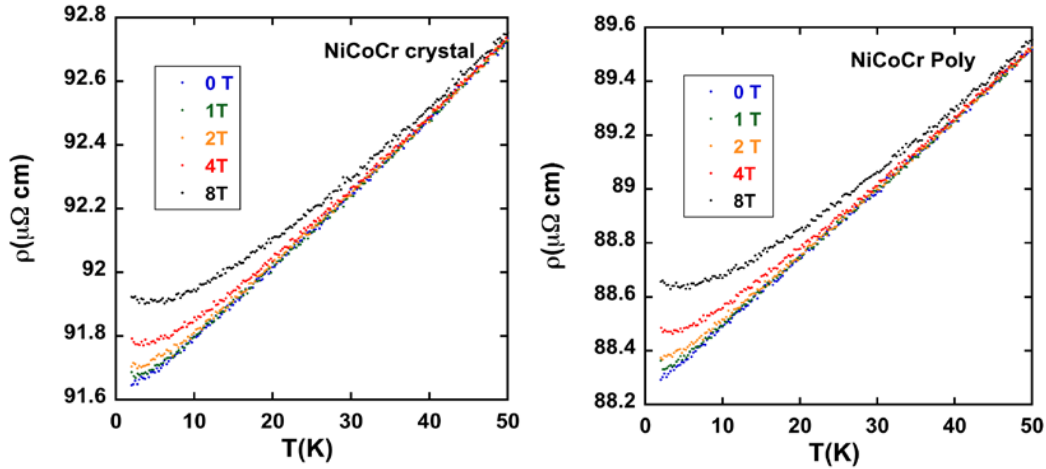

**Figure S3.** Resistivity of a NiCoCr single crystal (left) and a polycrystalline NiCoCr sample(right). The magnetic field is applied perpendicular to the current. The absolute values of the resistivity are only accurate to about 5% with most of the error determined by the measured distance between the voltage leads and the measured cross section area perpendicular to the current.

## S5 Magnetoresistance Data for other Cr Concentrations

The low temperature tranverse magnetoresistance data from of *all* of the NiCoCr$_x$ alloys are close to linear in **H**, even for samples with compositions outside of the critical region. Since there is no indication of linear Dirac bands from electronic structure calculations (**see S7**), the linear magnetoresistance is likely due to the chemical disorder present in these alloys. [2,3]

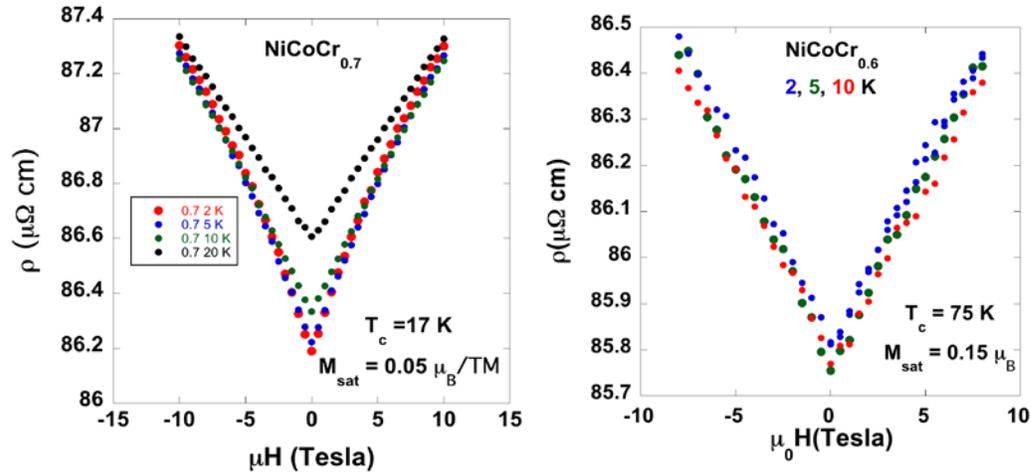

**Figure S4.** Transverse magnetoresistance for two of the weak itinerant ferromagnets NiCoCr$_{0.7}$ (left panel) and NiCoCr$_{0.6}$ (right panel)

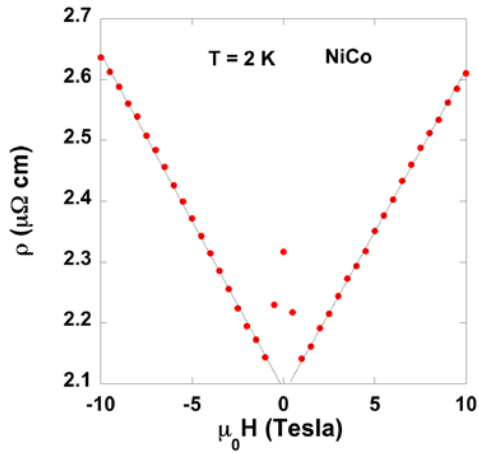

**Figure S5**. Transverse magnetoresistance for a NiCo single crystal (no Chromium) at 2 K. For low magnetic fields the resistance initially decreases as the ferromagnetic domains align, but for fields greater than 1 Tesla, the magnetoresistance is surprisingly linear.

## S6 Determination of Curie Temperature

The Curie temperature, $T_c$, is determined by Arrott plots ($M^2$ vs $H/M$) at fixed temperatures and by extrapolation of plots of $M^2$ vs T with $\mu_0 H = 0.01$ Tesla for temperatures below $T_c$.

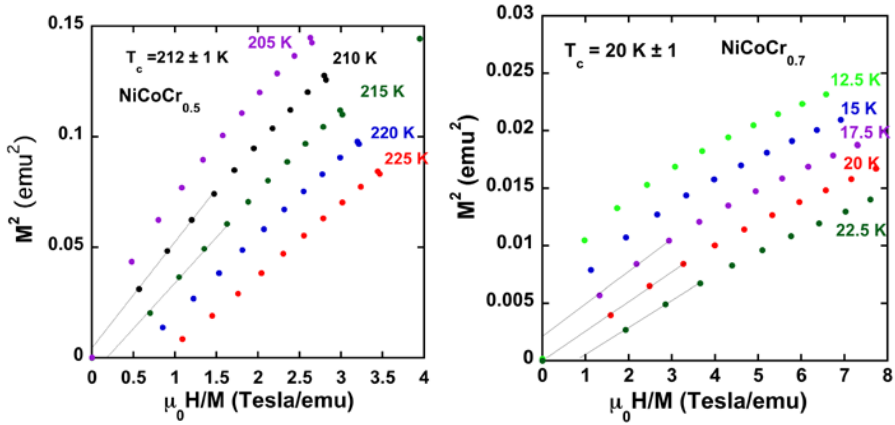

**Figure S6.** Arrott plot determination of Curie temperatures for NiCoCr$_{0.5}$ (left panel) and NiCoCr$_{0.7}$ (right panel).

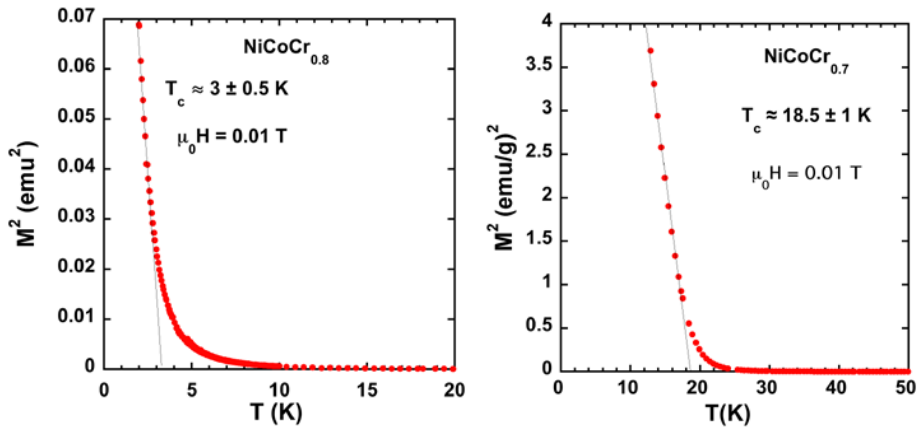

**Figure S7** Determination of Curie temperature for NiCoCr$_{0.8}$ (left panel) and NiCoCr$_{0.7}$ (right panel) from $M^2$ vs T fits just below $T_c$. (This assumes a mean field exponent)

## S6. Determination of $p_{eff}$ from Curie-Weiss plots well-above $T_c$ ($> \approx 2T_c$)

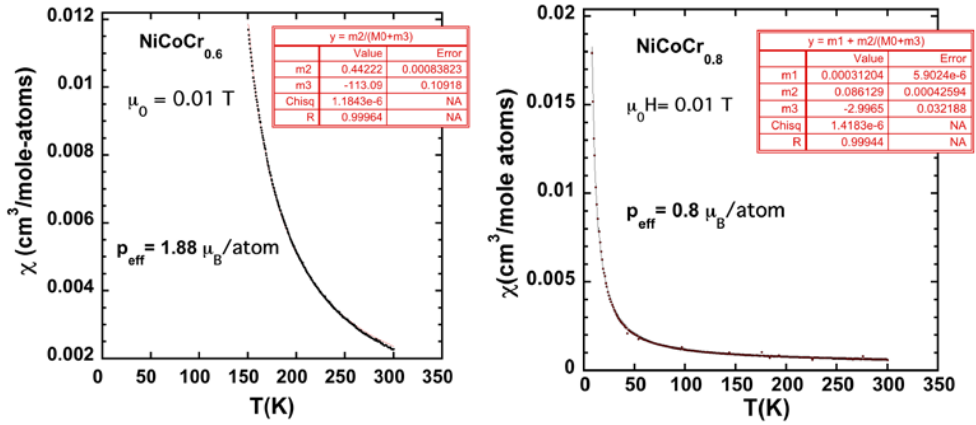

**Figure S8.** Curie-Weiss fits to NiCoCr$_{0.6}$ (left panel) and NiCoCr$_{0.8}$ (right panel) data at temperatures from 2T$_c$ to 300 K. As noted in the main text, for a weak itinerant ferromagnet the values of p$_{eff}$ should not be regarded as implying a large local magnetic moment on each atom but rather the values of p$_{eff}$ are determined by microscopic parameters of the band structure.

## S7. Mean field KKR-CPA calculations: methods and results

*Methods: Electronic structure calculations in substitutional alloys.*

The NiCoCr$_x$ electronic structure calculations were performed for an ideal fcc structure using the experimentally determined lattice parameters: 3.53 and 3.57 Å for NiCo and NiCoCr respectively. For the compositions x between 0 and 1 the lattice parameter was estimated according to Vegard's law. The residual resistivity was calculated using formalism based on one-electron Kubo formula[10-12] within fully relativistic Korringa-Kohn-Rostoker[4] Coherent-Potential-Approximation[5,6] (KKR-CPA) method, as implemented in SPRKKR code[7,8]. The KKR-CPA method describes the effects of chemical disorder on the electronic states, and delivers the configurationally averaged properties, for example, electronic structure, magnetic structure and transport properties. The calculations have been performed using the local spin density approximation LDA to density functional theory (DFT) and the Vosko, Wilk and Nusair parameterization of the exchange-correlation function[9]. The energy integration was executed over 32 points in complex energies plane. The Brillouin zone (BZ) summations over special k-points were over 28×28×28 k-points mesh during the density functional theory self-consistency cycle and 78×78×78 k-point mesh for the residual resistivity calculation. An angular momentum cutoff of 4 was used in the solution of the multiple-scattering equations.

Calculations of the configurationally averaged densities of states and Bloch spectral function along high-symmetry directions for pure nickel and the substitutionally disordered equiatomic NiCo and NiCoCr and NiCoFe alloys were performed using the ab initio KKR-CPA electronic structure method as implemented in the Hutsepot code. An angular momentum cutoff of 3 was used in the solution of the multiple-scattering equations. The KKR-CPA scattering-path matrix was calculated in reciprocal (k) space using a 20×20×20 k-point mesh during the density functional theory self-consistency cycle and 50×50×50 k-point mesh for the DOS calculation. The DOS and BSF were calculated at energy of 0.001 Rydberg off into the upper half of the complex plane, which gives rise to the slight broadening of the BSF for pure nickel.

*Results and discussion:*

To characterize the effects of chemical disorder on the electronic and magnetic properties, we performed electron structure calculations using ab initio Korringa–Kohn–Rostoker coherent-potential-approximation (KKR-CPA) method. The method, implemented within density functional theory, provides an ab initio theoretical description of the effects of disorder on the underlying electronic structure. An important aspect of the KKR-CPA is that it is specifically formulated to calculate the configurationally averaged electronic structure, including local and total densities of states (DOS) and magnetic moments within a single-site (or mean-field) theory. Moreover, the KKR-CPA has also been extended to calculate many other properties, including electron transport. Here it is worth noting that it is the configurationally averaged properties of random solid solutions that are actually measured by experiments. Figures S9, S10, and S11 show the Bloch spectral function (BSF), which is a generalization of the band structure of an ordered system to include disorder, and configurationally averaged DOS of NiCo, NiCoFe and NiCoCr, respectively. In these figures, the left and right-hand panels show the contributions from the spin-down (minority) and spin-up (majority) electrons the BSF and DOS, respectively. As can be seen from the BSF plots, for NiCo (Fig. S9) and NiCoFe (Fig. S10) the d-band smearing is largely limited to minority states; whereas in NiCoCr (Fig. S11), both minority and majority states are smeared, giving rise to the much larger overall smearing. The Fermi energy wave vector broadening of the BSF is related to the inverse of the electron mean free path. At the Fermi energy, k-space smearing implies a decrease in electron mean free path. No (or little) smearing implies an infinite (long) mean free path, whereas uniform smearing throughout the Brillouin zone implies a mean free path on the order of the interatomic spacing. While the electron mean free path for the NiCo and NiCoFe alloys (Fig. S9, S10) is short for the minority spin electrons, it is large for the majority spin electrons thereby providing a short circuit and an overall low resistivity. On the contrary, for the NiCoCr alloy (Fig. S11), both channels are broadened, particularly near the Fermi energy, implying a short mean free path in both spin channels. The consequently strong electron–electron scattering can lead to a high residual resistivity. Our calculated residual resistivity results are in full agreement with this qualitative explanation (Table S1). Thus, the resistivity in NiCoCr$_x$ monotonically increase from 2.28 to 70.0 µΩcm with composition x increase from 0 to 1.1.

**Table S1**. Average moment in Bohr magnetons on Ni, Co, and Cr atoms in NiCoCr$_x$ alloys as a function of Cr content x, as determined from KKR-CPA calculations. The average moment per atom in each alloy is also shown, as is the calculated residual resistivity in µΩcm for selected alloys. The results are calculated using SPR-KKR code[7,8]. A negative magnetic moment for Cr indicates it is antiparallel to the Ni and Co moments.

| x | $M_{Ni}$ | $M_{Co}$ | $M_{Cr}$ | $M_{av}$ | $\rho_0$ |
|---|---|---|---|---|---|
| 0.0 | 0.62 | 1.59 | -1.69 | 1.11 | 2.28 |
| 0.1 | 0.53 | 1.49 | -1.17 | 0.91 | 41.91 |
| 0.3 | 0.41 | 1.34 | -0.53 | 0.69 | 56.39 |
| 0.4 | 0.35 | 1.25 | -0.40 | 0.60 | 59.90 |
| 0.5 | 0.31 | 1.17 | -0.32 | 0.53 | 62.59 |
| 0.7 | 0.22 | 0.98 | -0.21 | 0.39 | 66.23 |
| 0.8 | 0.18 | 0.88 | -0.18 | 0.33 | 67.46 |
| 0.9 | 0.15 | 0.78 | -0.14 | 0.27 | 68.45 |
| 1.0 | 0.11 | 0.61 | -0.10 | 0.20 | 69.50 |
| 1.1 | 0.09 | 0.52 | -0.08 | 0.16 | 70.00 |

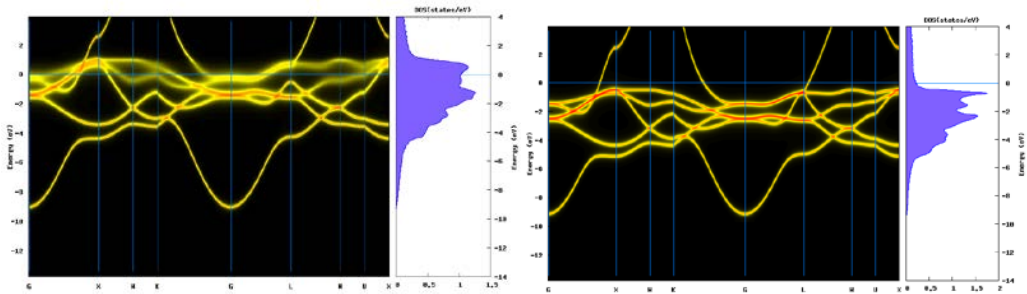

**Figure S9.** Band structure function and corresponding density of states NiCo minority (left) and majority (right) spin channel

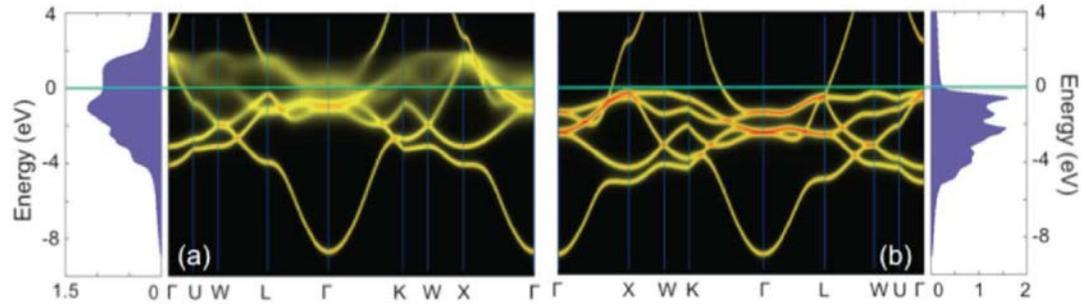

**Figure S10.** Band structure function and corresponding density of states in NiCoFe a) minority spin channel, b) majority spin channel.

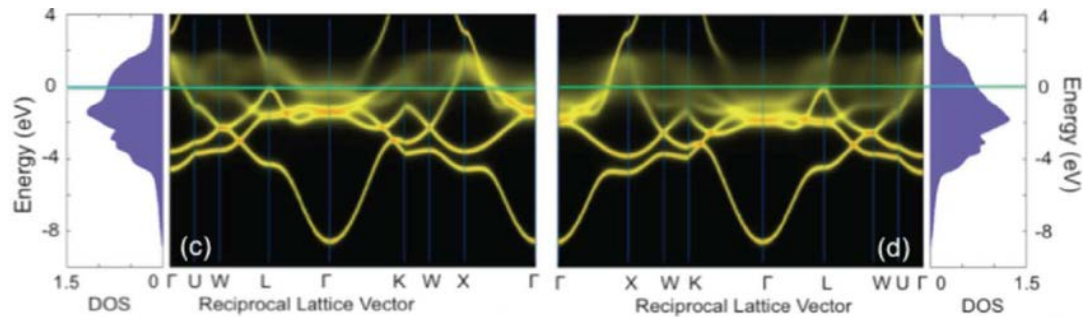

**Figure S11**. Band structure function and corresponding density of states in NiCoCr, c) minority spin channel, d) majority spin channel.

## References

1. Sales, B. C., Jin, R.Y., Mandrus, D., Orientation dependence of the anomalous Hall resistivity in single crystals of $Yb_{14}MnSb_{11}$, *Phys. Rev. B.* **77**, 024409-1-9 (2008).

2. Xu, R., Husmann, A., Rosenbaum, T. F., Saboungi, M. L., Enderby, J. E., Littlewood, P. B., Large magnetoresistance in non-magnetic silver chalcogenides. *Nature* **390**, 57-60 (1997).

3. Song, J. C. W., Refael, G., Lee P. A., Linear magnetoresistance in metals: Guiding center diffusion in a smooth random potential. *Phys. Rev. B.* **92**, 180204-1-5 (2015).

4. Korringa, J., On the calculation of the energy of a Bloch wave in a metal, Physica 13, 392 (1947). Kohn, W., Rostoker, N., Solution of the Schrödinger Equation in Periodic Lattices with an Application to Metallic Lithium, Phys.Rev. **94**, 1111 (1954).

5. Velický, B., Kirkpatrick, S., and Ehrenreich, H., Single-Site Approximations in the Electronic Theory of Simple Binary Alloys, Phys. Rev. **175**, 747 (1968).